\documentclass[a4paper,11pt]{article}
\pdfoutput=1 
\usepackage{jheppub} 
\usepackage[T1]{fontenc} 
\usepackage{amsmath,amssymb,epsfig,relsize,slashed,comment}
\usepackage[utf8]{inputenc}
\usepackage{ulem}
\allowdisplaybreaks

\definecolor{nicered}{rgb}{0.7,0.1,0.1}
\definecolor{nicegreen}{rgb}{0.1,0.5,0.1}
\definecolor{niceblue}{rgb}{0.0,0.1,0.7}
\hypersetup{colorlinks,citecolor=nicegreen,linkcolor=nicered,urlcolor=niceblue}

\def \beq{\begin{equation}}
\def \eeq{\end{equation}}
\def \bea{\begin{eqnarray}}
\def \eea{\end{eqnarray}}

\usepackage{amsmath}
\usepackage{amsfonts}
\usepackage{amssymb}
\usepackage{graphicx}
\usepackage{mathtools}
\usepackage{braket}
\usepackage{enumerate}
\usepackage{bbold}
\usepackage{slashed}
\usepackage{feynmp-auto}
\usepackage{subfigure}

\title{Singlet night in Feynman-ville: \\  one-loop matching of a real scalar}

\author[1]{Ulrich Haisch,}
\author[2]{Maximilian Ruhdorfer,}
\author[3]{Ennio Salvioni,}
\author[2]{\\ Elena Venturini,}
\author[2]{and Andreas Weiler$\,$}

\affiliation[1]{Max Planck Institute for Physics, F{\"o}hringer Ring 6,  80805 M{\"u}nchen, Germany}   
\affiliation[2]{Technische Universit{\"a}t M{\"u}nchen, Physik-Department, 85748 Garching, Germany}
\affiliation[3]{CERN, Theoretical Physics Department, 1211 Geneva 23, Switzerland}

\emailAdd{haisch@mpp.mpg.de}
\emailAdd{max.ruhdorfer@tum.de}
\emailAdd{ennio.salvioni@cern.ch}
\emailAdd{elena.venturini@tum.de}
\emailAdd{andreas.weiler@tum.de}

\abstract{\hspace{0.8mm}A complete one-loop matching calculation for real singlet scalar extensions of the Standard Model to the Standard Model effective field theory~(SMEFT) of dimension-six operators is presented. We compare our analytic results obtained by using Feynman diagrams to the expressions derived in the literature by a combination of the universal one-loop effective action (UOLEA) approach and Feynman calculus. After identifying contributions that have been overlooked in the existing calculations, we find that the pure diagrammatic approach and the mixed method lead to identical results. We highlight some of the subtleties involved in computing  one-loop matching corrections in SMEFT.}

\preprint{CERN-TH-2020-038\\
\hspace*{\fill}TUM-HEP-1254-20}

\begin{document} 

\maketitle

\section{Motivation}
\label{sec:introduction}

The Standard Model~(SM) effective field theory aka SMEFT provides a well-defined model-independent framework to characterise and to constrain new physics that is too heavy to be directly produced in laboratories. This virtue together with the lack of a clear evidence for direct production of new particles at the Large Hadron Collider~(LHC) has prompted considerable theoretical and experimental activities that led to the development of the SMEFT framework and its consistent and systematic application to LHC data. 

Important steps in the theoretical development of SMEFT were the classification of all independent dimension-six  SMEFT operators in~\cite{Grzadkowski:2010es}~(see also~\cite{Buchmuller:1985jz} for related earlier work) and the calculation of the full one-loop anomalous dimension matrix of these operators in a series of papers~\cite{Jenkins:2013zja,Jenkins:2013wua,Alonso:2013hga,Alonso:2014zka} --- partial one-loop and two-loop results have also been obtained in~\cite{Grojean:2013kd,Elias-Miro:2013mua,Zhang:2014rja,Pruna:2014asa,Brod:2014hsa,Cheung:2015aba,Gorbahn:2016uoy}. Considerable progress has also been made recently in improving the precision of matching calculations in perturbative extensions of the SM. A complete tree-level dictionary that allows to read off the  Wilson coefficients of the dimension-six SMEFT operators in any ultraviolet~(UV) completion with general scalar, spinor and vector field content and arbitrary interactions has been presented in~\cite{deBlas:2017xtg}. The computation of one-loop matching contributions has been advanced as well by the development of the so-called  universal one-loop effective action~(UOLEA) approach~\cite{Drozd:2015rsp} that generalised methods based on a covariant derivative expansion~(CDE)~\cite{Henning:2014wua}~(cf.~also~\cite{Gaillard:1985uh,Chan:1986jq,Cheyette:1987qz} for earlier works on functional techniques). In its initial formulation the UOLEA did not allow to compute the quantum effects associated to loops involving both heavy and light particles~\cite{delAguila:2016zcb}~(see also \cite{Boggia:2016asg}). This shortfall triggered several theoretical improvements aimed at capturing contributions of this type~\cite{Henning:2016lyp,Ellis:2016enq,Fuentes-Martin:2016uol,Zhang:2016pja,Ellis:2017jns,Kramer:2019fwz}. Despite the latter efforts the UOLEA formalism still remains incomplete to date, because a master formula that allows to calculate heavy-light contributions with open derivatives and mixed statistics has so far not been derived in the literature. However, see~\cite{Cohen:2019btp} for recent progress in this direction.

Computations in the UOLEA framework of the complete set of Wilson coefficients arising at the one-loop level in SMEFT therefore have to make use, at least partially, of conventional  Feynman diagram techniques. The article~\cite{Jiang:2018pbd} for example has employed the~UOLEA master formulae of~\cite{Ellis:2017jns} in combination with  Feynman calculus to obtain the first complete one-loop matching corrections at dimension-six for  real singlet scalar extensions of the SM~(SSM). In this work, we repeat the calculation of~\cite{Jiang:2018pbd} from scratch, relying entirely on the use of Feynman diagrams.  With the help of our independent computation we are able to identify terms that have been missed in the existing calculations~\cite{Ellis:2017jns,Jiang:2018pbd} --- small discrepancies in the latter publications have already been noted in the presentation~\cite{matchmaker}, but updated results have not been published so far. We stress that these discrepancies are due to oversights, and not due to structural limitations of the UOLEA approach. See~Appendix~\ref{sec:UOLEA} for further explanations. 

The results presented in this article constitute an important stepping stone to our forthcoming study~\cite{inprep} of the indirect collider sensitivity to pseudo Nambu-Goldstone boson~(pNGB) dark matter~(DM). This type of DM candidate is characterised by a parametric suppression of its scattering rate on ordinary matter, making it naturally compatible with the null results of direct DM searches. Strong theoretical motivations for this scenario come from composite Higgs models where the Higgs and DM both emerge as pNGBs~\cite{Frigerio:2012uc,Balkin:2018tma}, but the same DM phenomenology can also be realised in simple scalar extensions of the~SM~---~see~e.g.~\cite{Barger:2008jx,Gross:2017dan}. The collider reach on pNGB DM through tree-level production in vector boson fusion has been recently analysed~\cite{Ruhdorfer:2019utl}, finding a limited sensitivity even at future accelerators. This prompts us to explore one-loop probes, such as for instance off-shell Higgs production, employing an effective field theory (EFT) for the SM plus the~DM candidate~\cite{inprep}. A non-trivial aspect of such an analysis is that the virtual DM effects arise at the same order as those of DM-less, one-loop effective operators induced by heavy new physics. To clearly understand  the role of these DM-independent effects, it is useful to make use of an explicit model such as that of~\cite{Gross:2017dan}, integrating out the scalar radial mode to obtain precisely the dimension-six SMEFT operators considered in this work. This case study is especially useful because by varying the strength of the parameters, one can effectively interpolate between elementary Higgs-like and strongly-interacting light Higgs-like~\cite{Giudice:2007fh} EFT power countings within a simple setup. Given  its relevance to our upcoming work~\cite{inprep} and in view of the arguments presented in the previous paragraph, we believe it is worthwhile to provide the complete one-loop matching corrections for the SSM in this short note. 

This paper is organised as follows. In Section~\ref{sec:preliminaries} we specify our notation and conventions, while Section~\ref{sec:calculation}  contains the analytic results of the tree-level and one-loop matching calculation for the SSM. In Appendix~\ref{sec:UOLEA} we show  that the analytic results for the heavy one-loop matching corrections presented in Section~\ref{sec:oneloopmatching} can also be obtained in the UOLEA framework, while in Appendix~\ref{sec:SSMRG} we collect the anomalous dimensions that describe the renormalisation group~(RG) evolution of the SSM parameters and discuss their impact on our one-loop calculations. 

\section{Preliminaries}
\label{sec:preliminaries}

In order to set up our notation and conventions, let us first define the electroweak~(EW) part of the SM. Before spontaneous EW symmetry breaking the tree-level EW SM Lagrangian takes the following familiar form,
\bea
\begin{split} \label{eq:lagrangianSM}
\mathcal{L}_{\rm SM} & = ( D_\mu H)^\dagger (  D^\mu H   ) + \mu_h^2 |H|^2 - \frac12 \lambda_h |H|^4 - \frac14 B_{\mu\nu} B^{\mu\nu} - \frac14 W^a_{\mu\nu} W^{a\, \mu\nu} \\[1mm]
&  \phantom{xx} + \sum_{f\,=\,q,u,d,\ell,e} \bar{f} i \slashed{D} f  - \left(y_u \bar q \widetilde H u + y_d  \bar q H d+ y_e \bar \ell H e + \mathrm{h.c.} \right) \,.
\end{split}
\eea
Here $H$ denotes the SM Higgs doublet and the shorthand notation $\widetilde H_i = \epsilon_{ij} (H_j)^\ast$ with $\epsilon_{ij}$ totally antisymmetric and $\epsilon_{12}=1$ has been used. The covariant derivative is defined as 
\beq \label{eq:covariantderivative}
D_\mu = \partial_\mu - i g_1 Y B_\mu - i g_2 \hspace{0.25mm} \frac{\sigma^a}{2} \hspace{0.25mm}  W_\mu^a \,, 
\eeq
with $g_1$ and $g_2$ the $U(1)_Y$ and $SU(2)_L$ gauge coupling, respectively, and $B_\mu$ and $W_\mu^a$ ($B_{\mu \nu}$ and $W_{\mu \nu}^a$) the corresponding gauge fields (field strength tensors). The hypercharge operator is denoted by $Y$ with eigenvalues $\{Y_H, Y_q, Y_u, Y_d, Y_\ell, Y_e\} = \{1/2, 1/6, 2/3,-1/3,-1/2,-1\} $ and  $ \sigma^a$ are the Pauli matrices. The Yukawa couplings $y_u$, $y_d$ and $y_e$ are matrices in flavour space and a sum over flavour indices is implicit in (\ref{eq:lagrangianSM}). Finally, the symbols~$q$ and~$\ell$ denote left-handed quark and lepton doublets, while $u$, $d$ and $e$ are right-handed fermion~singlets.

As stated before, the goal of this article is to calculate the complete matching corrections up to  one-loop order that arise in the SSM.  At the renormalisable level, Lorentz and gauge  invariance allow a real singlet scalar  to couple to the SM exclusively through~$|H|^2$, and as a result the Lagrangian relevant for the further discussion can be written as 
\beq \label{eq:lagrangianphi}
\mathcal{L}_\phi = \frac{1}{2}\left(\partial_\mu\phi\right)^2 - \frac{1}{2}M^2\phi^2 - A|H|^2\phi -\frac{1}{2}\kappa |H|^2 \phi^2 - \frac{1}{3!} \mu \phi^3 - \frac{1}{4!} \lambda_\phi \phi^4 \, .
\eeq
Here we have ignored a potential tadpole contribution, meaning that the field  $\phi$ in~(\ref{eq:lagrangianphi}) corresponds to the excitation around a possible non-zero vacuum expectation value. The parameters $M^2, A, \kappa, \mu$ and $\lambda_\phi$ appearing in~(\ref{eq:lagrangianphi}) are treated as independent in what follows. 

\section{Calculation}
\label{sec:calculation}

By integrating out the field $\phi$ that appears in the SSM Lagrangian 
\beq \label{eq:lagrangianSSM}
{\cal L}_{{\rm SSM}} = {\cal L}_{\rm SM} + {\cal L}_\phi \,,
\eeq
one can determine the Wilson coefficients~$C_k$  that multiply the operators $Q_k$ in  SMEFT 
\beq \label{eq:lagrangianSMEFT}  
{\cal L}_{\rm SMEFT} =  \sum_k C_k Q_k  \,,
\eeq
 order by order in perturbation theory by performing a loop expansion  
\beq  \label{eq:perturbationtheory}
C_k = C_k^{(0)} + \frac{C_k^{(1)}}{(4 \pi)^2} + \ldots \,, 
\eeq
where $C_k^{(0)}$ and $C_k^{(1)}$ denote the  tree-level and one-loop coefficients, respectively. The notation introduced in~(\ref{eq:perturbationtheory}) will also be used when expanding other quantities of interest. The full  set of dimension-six SMEFT operators has been presented in the so-called Warsaw basis in~\cite{Grzadkowski:2010es}. Up to the one-loop level, it  turns out that matching the theory described by the Lagrangian~(\ref{eq:lagrangianSSM}) to the SMEFT Lagrangian~(\ref{eq:lagrangianSMEFT}) generates non-zero Wilson coefficients for the following set of~18 effective operators: 
\begin{align} \label{eq:warsaw}
Q_{H \Box} & = |H|^2\Box |H|^2 \,, &  Q_{Hu} & = (H^\dagger i \overset{\leftrightarrow}{D}_\mu H)(\bar u \gamma^\mu u) \,, \nonumber \\
Q_H  & = |H|^6 \,, & Q_{Hd} & = (H^\dagger i \overset{\leftrightarrow}{D}_\mu H)(\bar d \gamma^\mu d) \,,  \nonumber  \\
Q_{H D}  & = (H^\dagger D_\mu H)^\ast(H^\dagger D^\mu H) \,, & Q_{Hud}  & = ( i\hspace{0.2mm} \widetilde{H}^\dagger D_\mu H ) (\bar{u}\gamma^\mu d) \,, \nonumber   \\
Q_{H B} & = |H|^2 B_{\mu\nu}B^{\mu\nu} \,, & Q_{He} & = (H^\dagger i \overset{\leftrightarrow}{D}_\mu H)(\bar e \gamma^\mu e) \,, \nonumber \\
Q_{H W} & = |H|^2 W^{a}_{\mu \nu} W^{a\,\mu\nu} \,, &  Q^{(1)}_{H q} & = (H^\dagger i \overset{\leftrightarrow}{D}_\mu H)(\bar q \gamma^\mu q) \,, \\
Q_{H W\! B} & = (H^\dagger \sigma^a H) W^{a}_{\mu\nu} B^{\mu\nu} \,, & Q^{(3)}_{H q} & = (H^\dagger i \overset{\leftrightarrow}{D} \ \!\!\!^{\; a}_\mu H) (\bar q \gamma^\mu \sigma^a q) \,,  \nonumber \\
Q_{uH} & = |H|^2(\bar q \widetilde{H} u ) \,, & Q^{(1)}_{H \ell} & = (H^\dagger i \overset{\leftrightarrow}{D}_\mu H)(\bar\ell\gamma^\mu \ell) \,, \nonumber \\[1mm]
Q_{dH} & = |H|^2(\bar q H d) \,,  & Q^{(3)}_{H \ell} & = (H^\dagger i \overset{\leftrightarrow}{D} \ \!\!\!^{\; a}_\mu H) (\bar\ell\gamma^\mu\sigma^a \ell) \,, \nonumber  \\[1mm]
Q_{eH} & = |H|^2(\bar \ell  H e ) \, & Q_{2y} & = \big | \bar{q}_j y_u u  \epsilon^{ji} +   \bar{d} y_d^\dagger q^i +  \bar{e} y_e^\dagger \ell^i  \big |^2 \overset{}{} \,.\nonumber
\end{align}
Here $\Box = \partial_\mu \partial^\mu$, $H^\dagger i \overset{\leftrightarrow}{D}_\mu H = i H^\dagger \big (D_\mu -  \overset{\leftarrow}{D}_\mu \big ) H$ and $H^\dagger i \overset{\leftrightarrow}{D} \ \!\!\!^{\; a}_\mu H = i H^\dagger \big (\sigma^a D_\mu -  \overset{\leftarrow}{D}_\mu \sigma^a \big ) H$. For the operators $Q_{\psi H}$ with $\psi = u,d,e\hspace{0.2mm}$, as well as for $Q_{Hud}\hspace{0.2mm}$, the sum of the hermitian conjugate in~\eqref{eq:lagrangianSMEFT} is understood.

The matching of~(\ref{eq:lagrangianSSM}) onto~(\ref{eq:lagrangianSMEFT}) can be performed using either Feynman diagrams or functional methods. In fact, the work~\cite{Ellis:2017jns} employed the UOLEA approach to calculate the heavy~(i.e.~only $\phi$ loops) and the heavy-light~(i.e.~loops with both $\phi$ and Higgs exchange) one-loop matching corrections for the Wilson coefficients $C_{H \Box}$ and $C_{H}$. Based on the results of that article, the paper~\cite{Jiang:2018pbd} then presented the complete one-loop matching corrections in the model described by~(\ref{eq:lagrangianSSM}), computing the missing heavy-light contributions involving a $\phi$ scalar and a gauge boson or a fermion, by means of traditional Feynman diagram techniques.\footnote{The correction to the operator $Q_{Hud}$ was not given in~\cite{Jiang:2018pbd}. It was also missed in previous versions of this paper.}

In contrast to~\cite{Ellis:2017jns,Jiang:2018pbd} our calculation of the  Wilson coefficients $C_k^{(0)}$ and $C_k^{(1)}$ relies on Feynman diagrams only, and therefore represents an independent cross-check of the results obtained earlier. To allow for a direct comparison with the  expressions given in the publications~\cite{Ellis:2017jns,Jiang:2018pbd}, we regularise UV divergences using dimensional regularisation~(DR) in $d = 4 - 2 \epsilon$ dimensions and renormalise the results  in the  $\overline{\rm MS}$ scheme supplemented by the renormalisation scale~$\mu_R$.  Infrared~(IR) divergences have also been regularised dimensionally. The matching corrections can therefore be found by simply Taylor expanding the corresponding scattering amplitudes  in powers of external momenta squared divided by $M^2$ before performing any loop integration. On the other hand, SMEFT loop graphs do not contribute to the matching, because after Taylor expansion of the integrands they involve only scaleless integrals which vanish in DR --- see e.g.~\cite{Bobeth:1999mk,Gambino:2001au} for  further technical details. The actual generation and computation of the off-shell amplitudes made use of the {\tt Mathematica} packages {\tt FeynArts}~\cite{Hahn:2000kx}, {\tt FeynRules}~\cite{Alloul:2013bka}, {\tt FormCalc}~\cite{Hahn:1998yk} and {\tt Package-X}~\cite{Patel:2015tea}, and part of the  one-loop matching corrections obtained by computer were also verified  with pen and~paper. 

\subsection{Tree-level results}
\label{sec:treelevelmatching}

\begin{figure}[!t]
\begin{center}
\includegraphics[width=0.8\textwidth]{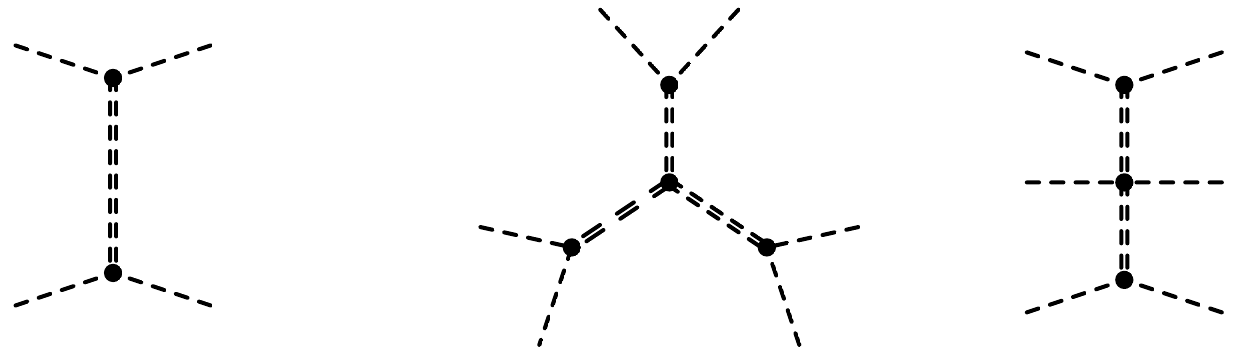} 
\vspace{2mm}
\caption{\label{fig:tree}  Example diagrams that contribute to the tree-level matching coefficients $\lambda_h^{(0)}$ and $C_{H \Box}^{(0)}$~(left) and $C_H^{(0)}$~(right), respectively. Double-dashed lines represent virtual exchange of the heavy real singlet scalar field $\phi$, while single-dashed lines  stand for $H$ or $H^\dagger$ fields.} 
\end{center}
\end{figure}

In the SM extension~(\ref{eq:lagrangianSSM}) only the Higgs quartic $|H|^4$~$\big($cf.~(\ref{eq:lagrangianSM})$\big)$ and the two effective operators  $Q_{H \Box}$  and $Q_H$~$\big($see~(\ref{eq:warsaw})$\big)$ receive a non-zero matching correction at tree level. The corresponding Feynman diagrams are shown in Figure~\ref{fig:tree}. For the additive shift $\lambda_h^{(0)}$ of the quartic Higgs coupling,~i.e.~$\lambda_h \to \lambda = \lambda_h + \lambda_h^{(0)}$, in agreement with~\cite{Jiang:2018pbd} we find 
\beq \label{eq:quartictree}
\lambda^{(0)}_h = -\frac{A^2}{M^2} \,,
\eeq
while in the case of the Wilson coefficients we obtain 
\begin{align}
C_{H \Box}^{(0)} & = -\frac{A^2}{2M^4} \,, \label{eq:CHBoxtree} \\[2mm]
C_{H}^{(0)} & = \frac{A^3 \mu}{6 M^6} -\frac{A^2 \kappa}{2M^4} \,. \label{eq:CHtree} 
\end{align}
The results~(\ref{eq:CHBoxtree}) and~(\ref{eq:CHtree}) are well-known and agree  with the analytic expressions reported for instance in the works~\cite{Henning:2014wua,deBlas:2017xtg,Jiang:2018pbd}.

\subsection{One-loop results}
\label{sec:oneloopmatching}

In order to determine the one-loop matching corrections $C_k^{(1)}$ to the Wilson coefficients of the dimension-six SMEFT operators $Q_k$ as given in~(\ref{eq:warsaw}), we consider only Feynman diagrams that are one-particle-irreducible in the light fields,~i.e.~we work  in the so-called Green's basis defined in~\cite{Jiang:2018pbd}, subsequently projecting our off-shell results onto the  Warsaw basis using the operator identities given in Appendix~A of the latter paper.

\begin{figure}[!t]
\begin{center}
\includegraphics[width=0.7 \textwidth]{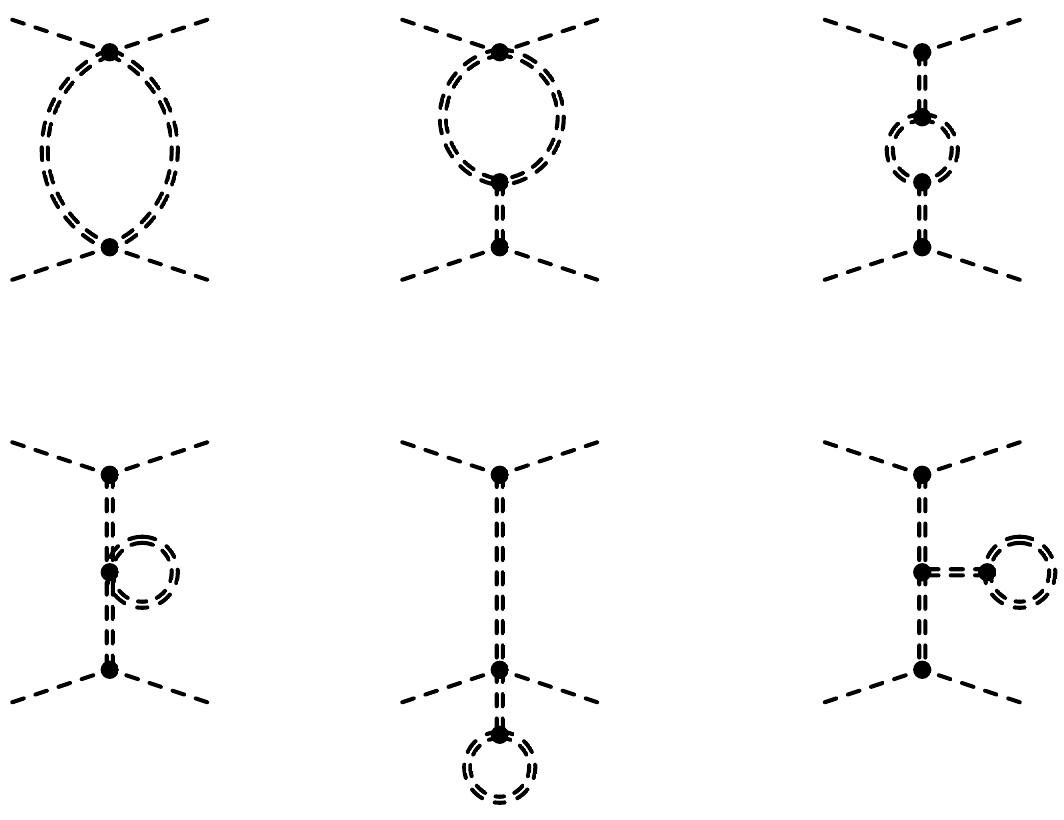} 
\vspace{0mm}
\caption{\label{fig:heavyhbox}  Examples of heavy contributions to the one-loop matching correction~$C_{H \Box}^{(1)}$. The line styles and their meanings resemble those of Figure~\ref{fig:tree}.} 
\end{center}
\end{figure}

The one-loop matching corrections of the tree-level operators $Q_{H \Box}$ and $Q_H$ receive contributions from three sources that we describe in the following. The first two types encode the threshold effects at a matching scale $\mu_M$ around $M$. The first kind of threshold corrections arise from heavy loops. For the case of $C_{H \Box}^{(1)}$, the relevant graphs are shown in Figure~\ref{fig:heavyhbox}. Notice that diagrams with $\phi$ tadpoles in general contribute to this type of corrections, hence the analytic expressions for the heavy contributions to the Wilson coefficients $C_{H \Box}^{(1)} $ and $C_H^{(1)}$ depend on how the tadpole contributions are fixed. In~our diagrammatic calculation, as well as in the UOLEA approach described in~Appendix~\ref{sec:UOLEA}, we renormalise $\phi$ tadpoles minimally~\cite{Jegerlehner:2001fb,Jegerlehner:2002em} and~(\ref{eq:CHBoxloop}), (\ref{eq:CHloop}), (\ref{eq:CHBox1heavy}) and~(\ref{eq:CH1heavy}) therefore correspond to the $\overline{\rm MS}$ scheme. Notice that  in the $\overline{\rm MS}$~scheme the effective one-loop scalar potential  contains a term linear in the $\phi$ field, which by definition would be absent in the on-shell scheme where the tadpole counterterm is fixed such that all tadpole diagrams vanish~\cite{Denner:1991kt}~---~see also~\cite{Fleischer:1980ub,Denner:2016etu,Cullen:2019nnr} for excellent discussions of the different treatments of tadpoles.

\begin{figure}[!t]
\begin{center}
\includegraphics[width=0.8 \textwidth]{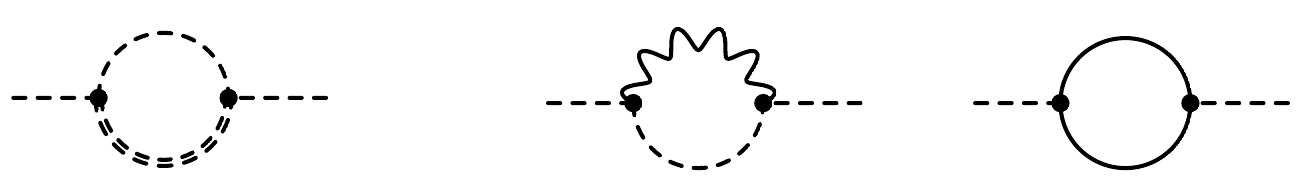} 
\vspace{2mm}
\caption{\label{fig:wfr} Higgs wave function renormalisation effects. Left: heavy-light contribution to $Z_H^{(1)}$ in~\eqref{eq:ZH1}. Right: gauge-boson and fermionic contributions to the anomalous dimensions of the SSM parameters $A$ and $\kappa$ in~\eqref{eq:gammmaA} and \eqref{eq:gammmakappa}. The wiggly line corresponds to a $B_\mu$ or $W_\mu^a$ field, the solid straight lines represent  fermion fields, while the rest of the line styles and their meanings are identical to those employed in~Figure~\ref{fig:tree}.}
\end{center}
\end{figure}

\begin{figure}[!t]
\begin{center}
\includegraphics[width=0.7 \textwidth]{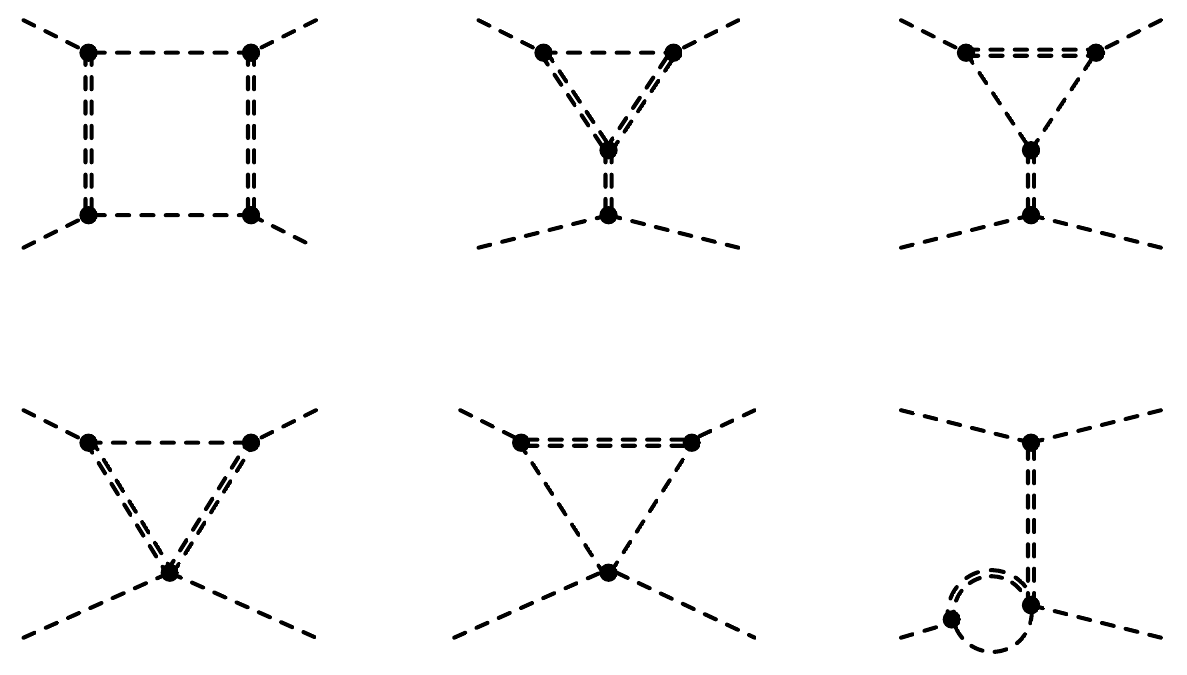} 
\vspace{2mm}
\caption{\label{fig:heavylighthbox}  Examples of heavy-light  scalar contributions to the one-loop matching correction~$C_{H \Box}^{(1)}$. Diagrams with loops that contain only $H$  fields are not shown, since they evaluate to zero in DR if the $H$ fields are taken to be massless before Taylor expanding the corresponding loop integrals. The line styles and their meanings resemble those of Figure~\ref{fig:tree}.} 
\end{center}
\end{figure}

\begin{figure}[!t]
\begin{center}
\includegraphics[width=0.2 \textwidth]{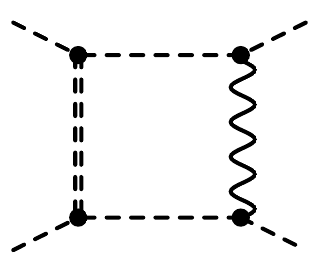}
\vspace{2mm}
\caption{\label{fig:heavylightgaugehbox} Example of a heavy-light  gauge-boson contribution to the one-loop matching correction~$C_{H \Box}^{(1)}$. Graphs with loops of only $H$ and $B_\mu$, or $H$ and $W_\mu^a$  fields are not displayed, because such diagrams do not contribute  if IR divergences are regulated dimensionally.  The line styles and their meanings mirror those  in~Figure~\ref{fig:wfr}.}
\end{center}
\end{figure}

The second type of threshold corrections to $C_{H \Box}^{(1)} $ and $C_H^{(1)}$ stem from heavy-light loop diagrams involving either a $H$ or a $B_\mu$~($W^a_\mu$) field. Universal effects related to the wave function renormalisation of the Higgs field belong to this class. In fact, after the field redefinition $H \to \big (1 -  Z_H^{(1)}/(4 \pi)^2 \big ) H$ the Wilson coefficients $C_{H \Box}$ and $C_H$ receive a one-loop contribution proportional to $C_{H \Box}^{(0)} $ and $C_H^{(0)}$, respectively. The relevant wave function renormalisation constant $Z_H^{(1)}$ is determined by calculating the one-loop corrections to the Higgs kinetic term $( D_\mu H)^\dagger (  D^\mu H   )$ that arises from the graph displayed on the left in Figure~\ref{fig:wfr}. In agreement with~\cite{Jiang:2018pbd} we obtain
\beq \label{eq:ZH1}
Z_H^{(1)} = \frac{A^2}{4M^2} \, .
\eeq
In addition to the wave function renormalisation contributions, non-universal heavy-light corrections arise. The corresponding scalar and gauge-boson contributions to $C_{H \Box}^{(1)}$ are displayed in Figure~\ref{fig:heavylighthbox} and Figure~\ref{fig:heavylightgaugehbox}, respectively.  Notice that when IR divergences are regulated dimensionally, SSM diagrams involving only light particles in the loop do not need to be considered, because such graphs result in scaleless integrals after Taylor expanding the associated off-shell amplitudes in powers of external momenta squared divided by $M^2$. This should be contrasted to methods that use small external momenta or small  light-field masses as IR regulators (cf.~for instance~\cite{Buras:1998raa,Gambino:1998rt,Buras:1999st}). In these cases, SSM diagrams with only light particles in the loop give non-zero IR divergent corrections but their contributions are exactly cancelled by the corresponding SMEFT graphs. As a result, the one-loop matching corrections $C_{H \Box}^{(1)}$ and $C_H^{(1)}$ turn out to be independent of the procedure that is used to regulate IR divergences (as they should), and in our calculation we have employed DR to regulate both UV and IR divergences simply because it is technically the easiest method to implement.

\begin{figure}[!t]
\begin{center}
\includegraphics[width=0.85 \textwidth]{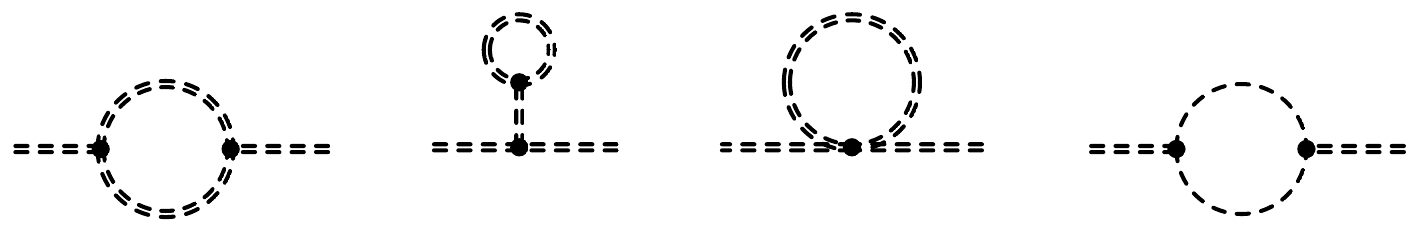}
\vspace{2mm}
\caption{\label{fig:phiprop}  One-loop contributions to the propagator of the real singlet scalar. In the $\overline{\rm MS}$ scheme the UV poles of the first and second diagram cancel against each other. The line styles and their meanings are identical to those employed in~Figure~\ref{fig:tree}.} 
\end{center}
\end{figure}

\begin{figure}[!t]
\begin{center}
\includegraphics[width=0.65 \textwidth]{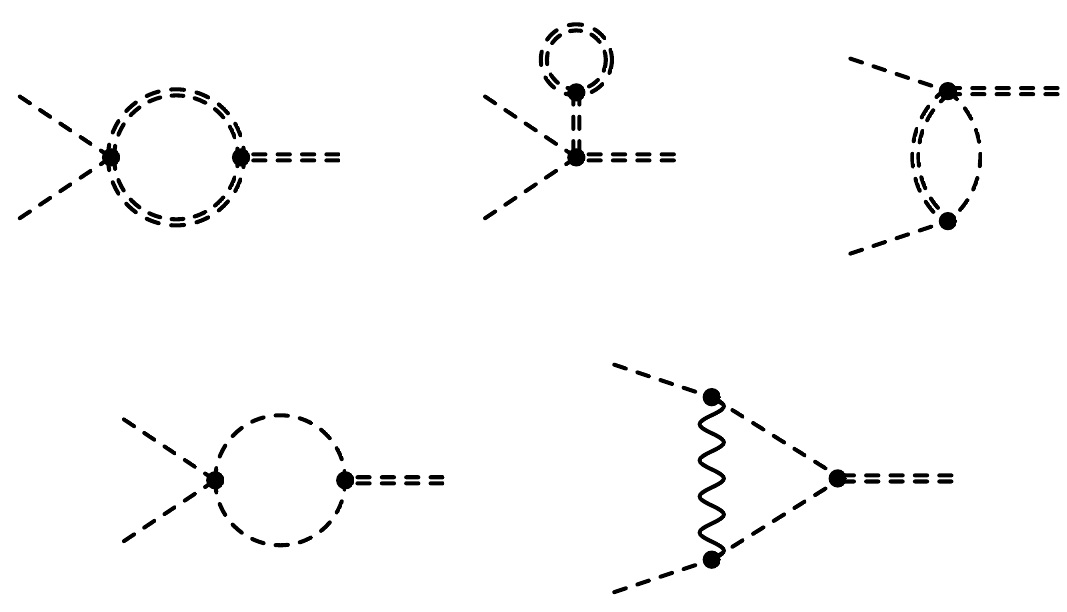}
\vspace{2mm}
\caption{\label{fig:A} Examples of one-loop contributions to the renormalisation of the coupling $A$. The UV poles of the first and second graph cancel against each other in the $\overline{\rm MS}$ scheme. The line styles and their meanings are analogue to those of Figure~\ref{fig:heavylightgaugehbox}.   } 
\end{center}
\end{figure}

\begin{figure}[!t]
\begin{center}
\includegraphics[width=0.9 \textwidth]{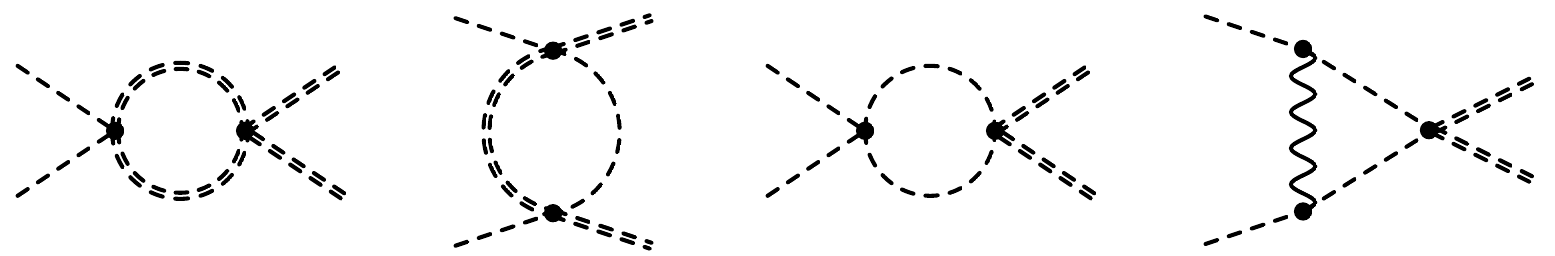}
\vspace{2mm}
\caption{\label{fig:kappa} Examples of one-loop contributions to the renormalisation of the coupling  $\kappa$. The line styles and their meanings are analogue to those of Figure~\ref{fig:heavylightgaugehbox}.  } 
\end{center}
\end{figure}

\begin{figure}[!t]
\begin{center}
\includegraphics[width=0.65 \textwidth]{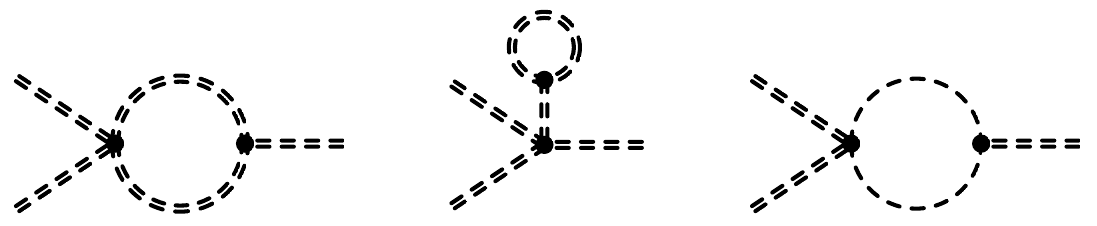} 
\vspace{2mm}
\caption{\label{fig:mu} Examples of one-loop contributions to the renormalisation of the coupling  $\mu$. The line styles and their meanings resemble those of Figure~\ref{fig:tree}. } 
\end{center}
\end{figure}

The third type of corrections to  $C_{H \Box}^{(1)} $ and $C_H^{(1)}$ arise instead from the renormalisation of the SSM parameters that enter the tree-level Wilson coefficients~(see e.g.~\cite{Cohen:2019wxr} for a pedagogical discussion). These contributions are, therefore, purely logarithmic in the $\overline{\rm MS}$ scheme. The logarithmic terms proportional to the SM couplings generate a RG flow that extends below the matching scale, providing one of the contributions to the anomalous dimensions of the SMEFT Wilson coefficients. These effects, coming from light loops, give the same corrections to both the SSM and SMEFT amplitudes. On the other hand, the RG-flow contributions proportional to UV SSM parameters appear only above~$\mu_M$. The anomalous dimension of~$M^2$ is obtained from the Feynman diagrams shown in Figure~\ref{fig:phiprop}, while Figure~\ref{fig:A},  Figure~\ref{fig:kappa} and Figure~\ref{fig:mu} display example graphs of contributions to the running of $A$, $\kappa$ and $\mu$, respectively.  Notice that the anomalous dimensions that describe the RG flow of the SSM parameters $A$ and $\kappa$ also contain pieces arising from the light contributions to the Higgs wave function displayed on the right-hand side in Figure~\ref{fig:wfr}, since the corresponding operators contain two powers of the $H$ field. See Appendix~\ref{sec:SSMRG} for further details and explanations.

In the case of the dimension-six SMEFT operator $Q_{H\Box}$, we find after combining the three different types of contributions described above the following result for the one-loop matching correction:
\beq 
\begin{split} \label{eq:CHBoxloop}
C_{H \Box}^{(1)} & = -\frac{\kappa ^2}{24 M^2}+\frac{25 A^2 \kappa -6 A^2 \lambda _{\phi }-5 A \kappa  \mu }{12 M^4} +\frac{38 A^4-26 A^3 \mu +11 A^2 \mu ^2}{24 M^6} \\[1mm] 
& \phantom{xx}  -\frac{31A^2 \left (g_1^2+3 g_2^2 \right )}{72 M^4} + \gamma_{H\Box, H\Box}  \hspace{0.5mm}  C_{H \Box}^{(0)}  \hspace{0.25mm} \ln \frac{\mu_M}{M}  \,.
\end{split}
\eeq
Here
\beq \label{eq:ADHBox}
 \gamma_{H\Box, H\Box} = 12 \lambda - \frac{4}{3} \left (g_1^2 + 3 g_2^2 \right)+ 4 y_2 \,, 
\eeq
with $\lambda$ denoting the quartic Higgs coupling that includes the tree-level shift~(\ref{eq:quartictree}) and the objects $C_{H \Box}^{(0)}$ and $y_2$ defined in~(\ref{eq:CHBoxtree}) and~(\ref{eq:y2}), respectively. Notice that all mass and coupling parameters in (\ref{eq:CHBoxloop}), (\ref{eq:ADHBox}) as well as in the tree-level expression \eqref{eq:CHBoxtree} are renormalised at the scale $M$.

A~couple of comments concerning our result~(\ref{eq:CHBoxloop}) seem to be in order. The rational terms in~(\ref{eq:CHBoxloop}) receive contributions from the Higgs wave function renormalisation constant~(\ref{eq:ZH1}) and heavy and heavy-light  diagrams (see~Figure~\ref{fig:heavyhbox}, Figure~\ref{fig:heavylighthbox} and Figure~\ref{fig:heavylightgaugehbox}), while the logarithmic terms result from the combination of heavy and heavy-light  graphs as well as the renormalisation of $M^2$ and~$A$ according to~(\ref{eq:gammmaM2}) and~(\ref{eq:gammmaA}). In fact, the logarithmic pieces proportional to SM couplings combine to give the anomalous dimension~$\gamma_{H\Box, H\Box}$, whereas the remaining terms cancel, because they do not run below the matching scale (see~Appendix~\ref{sec:SSMRG} for further details). The anomalous dimension describes the self-mixing of the dimension-six operator~$Q_{H \Box}$, and our expression~(\ref{eq:ADHBox}) agrees with the results of the direct calculation of~$\gamma_{H\Box, H\Box}$ presented in~\cite{Jenkins:2013zja,Jenkins:2013wua,Alonso:2013hga} --- the found agreement constitutes a non-trivial cross-check of our computation. We add that the logarithmic corrections in~(\ref{eq:CHBoxloop}) are scheme-independent, while the rational terms in  $C_{H \Box}^{(1)}$ depend on the choice of renormalisation scheme,  including the  specific treatment of $\phi$ tadpoles. Notice that the cancellation in~(\ref{eq:CHBoxloop}) of logarithms that are not proportional to SM couplings is crucial to achieve the correct factorisation of short-distance and long-distance effects. In fact, in the SMEFT only the combination of SSM parameters that forms a Wilson coefficient has a non-trivial RG flow, together with the SM couplings $\lambda$, $g_1$, $g_2$ and $y_f$. Thus, the correct description of long-distance physics has to be formulated in terms of the SM couplings and the Wilson coefficient~$C_{H \Box}$  evaluated at the low-energy scale. Let us finally mention that~(\ref{eq:CHBoxloop}) differs  from the expression for $C_{H \Box}^{(1)}$ given in both~\cite{Ellis:2017jns} and~\cite{Jiang:2018pbd}. The disagreement has two sources. First, as shown in~Appendix~\ref{sec:UOLEA}, the latter calculations miss certain heavy-loop contributions, and second, RG effects associated to the running of SSM parameters have not been explicitly included in the existing computations.

In the case of the operator $Q_H$,  we have calculated the $H \to H$, $HHH \to HHH$, $HH^\dagger \to W_\mu^a$ and $HH \to HH W_\mu^a W_\nu^b$  scattering amplitudes to find the following expression for the one-loop correction to the Wilson coefficient $C_H$:
\bea
\begin{split} \label{eq:CHloop} 
C_{H}^{(1)} & = -\frac{\kappa^3}{12 M^2} - \frac{6 A^2 \kappa  \lambda_\phi + 162 A^2 \kappa \lambda - 66 A^2 \kappa ^2 - 164 A^2  \lambda^2 + 3 A \kappa^2 \mu }{12 M^4} \\[1mm]
& \phantom{xx}  + \frac{87 A^4 \kappa - 6 A^4 \lambda_\phi - 72 A^4 \lambda - 60 A^3 \kappa \mu + 4 A^3 \mu \lambda_\phi +78 A^3 \lambda \mu + 6 A^2 \kappa \mu^2}{12 M^6}  \\[1mm]
& \phantom{xx}  -\frac{8 A^6  + 21 A^5 \mu - 12 A^4 \mu^2 + 2 A^3 \mu^3}{12 M^8} - \frac{31 A^2 \lambda g_2^2}{18 M^4}  \\[1mm]
& \phantom{xx} + \left (  \gamma_{H,H\Box} \hspace{0.25mm} C_{H \Box}^{(0)}  + \gamma_{H,H} \hspace{0.25mm} C_H^{(0)} \right ) \ln \frac{\mu_M}{M} \,.
\end{split}
\eea
The anomalous dimensions entering~(\ref{eq:CHloop}) read 
\begin{align}
\gamma_{H,H\Box} &= -40 \lambda^2 + \frac{20 \lambda g_2^2}{3} \,, \label{eq:gammaHHBox} \\[1mm]
\gamma_{H,H} & = 54 \lambda - \frac{9}{2} \left (g_1^2 + 3 g_2^2 \right ) + 6 y_2 \label{eq:gammaHH}  \,, 
\end{align}
while~(\ref{eq:CHBoxtree}), (\ref{eq:CHtree}) and (\ref{eq:y2}) contain  the explicit expressions for $C_{H \Box}^{(0)}$, $C_H^{(0)}$ and $y_2$. All mass and coupling parameters that appear in (\ref{eq:CHloop}) to (\ref{eq:gammaHH}) as well as in the tree-level expressions~\eqref{eq:CHBoxtree} and~\eqref{eq:CHtree} are renormalised at the scale $M$.

\begin{figure}[!t]
\begin{center}
\includegraphics[width=0.85 \textwidth]{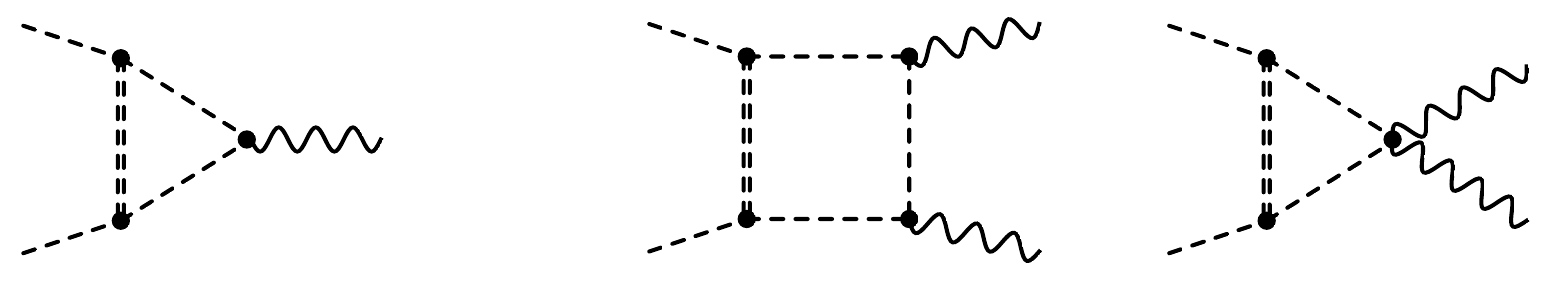}
\vspace{2mm}
\caption{\label{fig:heavylighthhgb} Examples of one-loop heavy-light scalar diagrams that need to be considered to extract the one-loop matching corrections of the bosonic dimension-six SMEFT operators in~(\ref{eq:warsaw}) that do not receive a tree-level Wilson coefficient. The line styles and their meanings resemble those of Figure~\ref{fig:heavylightgaugehbox}.} 
\end{center}
\end{figure}

Like in the case of~(\ref{eq:CHBoxloop}), one observes that the logarithmic corrections in~(\ref{eq:CHloop}) involve only anomalous dimensions that depend on SM couplings, but not on SSM parameters. In fact, our expressions~(\ref{eq:gammaHHBox}) and~(\ref{eq:gammaHH}) for~$\gamma_{H,H\Box}$ and $ \gamma_{H,H}$ agree with the results obtained in the articles~\cite{Jenkins:2013zja,Jenkins:2013wua,Alonso:2013hga}.  The source of the  difference between the first four terms in~(\ref{eq:CHloop}) and the rational terms of $C_{H}^{(1)}$ as quoted in \cite{Ellis:2017jns,Jiang:2018pbd} is unraveled in Appendix~\ref{sec:UOLEA}. In addition, the existing calculations do not explicitly include effects stemming from the renormalisation of SSM parameters~---~cf.~(\ref{eq:gammmaM2}) to (\ref{eq:gammmamu}) --- and therefore the logarithmic corrections given in~(\ref{eq:CHloop}) differ from the corresponding terms specified in~\cite{Ellis:2017jns,Jiang:2018pbd}  as well.

In the case of the 16 dimension-six SMEFT operators in~(\ref{eq:warsaw}) that do not receive a tree-level Wilson coefficient, only heavy-light Feynman diagrams contribute to the one-loop matching. To extract the relevant one-loop  corrections of the bosonic operators, we have calculated the off-shell amplitudes for $H H \to HH$, $H H^\dagger \to V_\mu$ and $H H^\dagger \to V_\mu V_\nu^\prime$ scattering with $V^{(\prime)}_\mu = B_\mu, W^a_\mu$. See~Figure~\ref{fig:heavylighthhgb} for the processes with external gauge bosons. We obtain
\begin{align}
C_{HD}^{(1)}  & = -\frac{31 A^2 g_1^2}{18 M^4} + \gamma_{HD,H\Box} \hspace{0.5mm} C_{H\Box}^{(0)}  \hspace{0.5mm}  \ln \frac{\mu_M}{M}  \,,  \label{eq:CHDloop} \\[1mm]
C_{HB}^{(1)}  &= \frac{A^2 g_1^2}{12 M^4}  \,,   \label{eq:CHBloop}  \\[1mm]
C_{HW}^{(1)}  &=  \frac{A^2 g_2^2 }{12 M^4}  \,,  \label{eq:CHWloop}  \\[1mm]
C_{HWB}^{(1)}  &=   \frac{A^2 g_1 g_2}{6 M^4}  \label{eq:CHWBloop}  \,. 
\end{align}
Here 
\beq \label{eq:gammaHDHBox}
\gamma_{HD,H\Box}  = \frac{20 g_1^2}{3} \,,
\eeq
and the expression for the tree-level Wilson coefficient $C_{H\Box}^{(0)} $ has already been given in~(\ref{eq:CHBoxtree}). We emphasise that our results~(\ref{eq:CHDloop}) to~(\ref{eq:CHWBloop}) agree with~(A.20) to (A.23)~of~\cite{Jiang:2018pbd} and that the anomalous dimension~(\ref{eq:gammaHDHBox}) matches that calculated in~\cite{Alonso:2013hga}. Notice that in contrast to~$C_{HD}^{(1)}$, the Wilson coefficients $C_{HB}^{(1)}$, $C_{HW}^{(1)}$ and  $C_{HWB}^{(1)}$ do not receive logarithmic corrections. This feature is expected, because the tree-level operators $Q_{H \Box}$ and $Q_H$ do not mix into $Q_{HB}$, $Q_{HW}$ and $Q_{HWB}$ at the one-loop level~\cite{Jenkins:2013zja,Jenkins:2013wua,Alonso:2013hga}.

\begin{figure}[!t]
\begin{center}
\includegraphics[width=0.55 \textwidth]{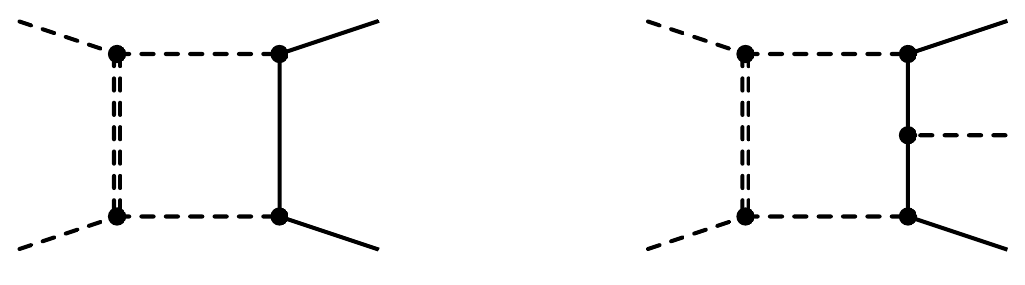}
\vspace{2mm}
\caption{\label{fig:heavylighthhf} Examples of one-loop heavy-light scalar diagrams that need to be considered to extract the one-loop matching corrections of the fermionic dimension-six SMEFT operators in~(\ref{eq:warsaw}).  The line styles and their meanings duplicate those of Figure~\ref{fig:wfr}.}
\end{center}
\end{figure}

In order to determine the one-loop matching corrections of the fermionic dimension-six SMEFT operators appearing in~(\ref{eq:warsaw}), we have computed the heavy-light scalar contributions to the $H H^\dagger \to \bar f f$ and $H H \to H \bar f f$ off-shell amplitudes with $f = q, u, d, \ell, e$, as well as to $HH\to u\bar{d}$. Examples of the corresponding diagrams are shown in Figure~\ref{fig:heavylighthhf}.  We find 
\begin{align}
C^{(1)}_{\psi H} & = -\frac{A^2 y_\psi }{36 M^4} \left ( 27 \kappa - 87 \lambda - \frac{9 A \mu}{M^2} + 31 g_2^2 - 45 y_\psi^\dagger y_\psi \right ) + \gamma_{\psi H, H\Box} \hspace{0.5mm} C_{H \Box}^{(0)}  \hspace{0.5mm} \ln \frac{\mu_M}{M} \,, \label{eq:C1fH} \\[1mm]
C^{(1)}_{Hu} & = -\frac{A^2}{216 M^4}  \left ( 34 g_1^2 - 135 y_u^\dagger y_u \right ) + \gamma_{Hu, H\Box} \hspace{0.5mm} C_{H \Box}^{(0)}  \hspace{0.5mm} \ln \frac{\mu_M}{M} \,, \label{eq:C1Hu} \\[1mm]
C^{(1)}_{Hd} & = \frac{A^2}{216 M^4}  \left ( 17 g_1^2 - 135 y_d^\dagger y_d \right ) + \gamma_{Hd, H\Box} \hspace{0.5mm} C_{H \Box}^{(0)}  \hspace{0.5mm} \ln \frac{\mu_M}{M} \,, \label{eq:C1Hd} \\[1mm]
C^{(1)}_{Hud}  & = -\frac{5 A^2}{4 M^4}\, y_u^\dagger y_d  +  \gamma_{Hud, H\Box} \hspace{0.5mm} C_{H\Box}^{(0)}  \hspace{0.5mm} \ln \frac{\mu_M}{M} \,,  \label{eq:C1Hud} \\[1mm]
C^{(1)}_{He} & = \frac{A^2}{72 M^4}  \left ( 17 g_1^2 - 45 y_e^\dagger y_e \right ) + \gamma_{He, H\Box} \hspace{0.5mm} C_{H \Box}^{(0)} \hspace{0.5mm} \ln \frac{\mu_M}{M} \,, \label{eq:C1He} \\[1mm]
C^{(1)}_{{Hq}^{(1)}} & = -\frac{A^2}{432 M^4} \left [ 17 g_1^2 + 135 \left ( y_u y_u^\dagger - y_d y_d^\dagger \right ) \right ] +  \gamma_{Hq^{(1)}, H\Box} \hspace{0.5mm} C_{H\Box}^{(0)}  \hspace{0.5mm} \ln \frac{\mu_M}{M} \,,  \label{eq:C1Hq1} \\[1mm]
C^{(1)}_{{Hq}^{(3)}} & = -\frac{A^2}{144 M^4} \left [ 17 g_2^2 - 45 \left ( y_u y_u^\dagger + y_d y_d^\dagger \right ) \right ] +  \gamma_{Hq^{(3)}, H\Box} \hspace{0.5mm} C_{H\Box}^{(0)}  \hspace{0.5mm} \ln \frac{\mu_M}{M} \,,  \label{eq:C1Hq3} \\[1mm]
C^{(1)}_{{H\ell}^{(1)}} & = \frac{A^2}{144 M^4} \left (17 g_1^2 + 45 y_e y_e^\dagger \right )  +  \gamma_{H\ell^{(1)}, H\Box} \hspace{0.5mm} C_{H\Box}^{(0)}  \hspace{0.5mm} \ln \frac{\mu_M}{M} \,,  \label{eq:C1Hl1} \\[1mm]
C^{(1)}_{{H\ell}^{(3)}} & = -\frac{A^2}{144 M^4} \left ( 17 g_2^2 - 45 y_e y_e^\dagger \right )+  \gamma_{H\ell^{(3)}, H\Box} \hspace{0.5mm} C_{H\Box}^{(0)}  \hspace{0.5mm} \ln \frac{\mu_M}{M} \,,  \label{eq:C1Hl3} \\[1mm]
C^{(1)}_{2y} & = \frac{A^2}{6M^4} \,.  \label{eq:C12y}
\end{align} 
The one-loop anomalous dimensions appearing in the above expressions are 
\begin{align}
\gamma_{\psi H, H\Box} & = -y_\psi \left ( 2 \lambda - \frac{10 g_2^2}{3} + 6 y_\psi^\dagger y_\psi \right ) \,, \label{eq:ADC1fH} \\[1mm]
\gamma_{Hu, H\Box} & =  \frac{2 g_1^2}{9} -  y_u^\dagger y_u  \,, \label{eq:ADC1Hu} \\[1mm]
\gamma_{Hd, H\Box} & =  -\frac{g_1^2}{9} +  y_d^\dagger y_d  \,, \label{eq:ADC1Hd} \\[1mm]
\gamma_{Hud, H\Box}  & =  2\hspace{0.1mm} y_u^{\dagger} y_d  \,,  \label{eq:ADC1Hud}\\[1mm]
\gamma_{He, H\Box} & =  -\frac{g_1^2}{3} +  y_e^\dagger y_e  \,, \label{eq:ADC1He} \\[1mm]
\gamma_{Hq^{(1)}, H\Box} & =  \frac{g_1^2}{18}  + \frac{1}{2} \left (  y_u y_u^\dagger - y_d y_d^\dagger \right )  \,,  \label{eq:ADC1Hq1} \\[1mm]
\gamma_{Hq^{(3)}, H\Box} & =  \frac{g_2^2}{6}  - \frac{1}{2} \left (  y_u y_u^\dagger + y_d y_d^\dagger \right )  \,,  \label{eq:ADC1Hq3} \\[1mm]
\gamma_{H\ell^{(1)}, H\Box} & =  -\frac{g_1^2}{6}  - \frac{1}{2}  \hspace{0.25mm} y_e y_e^\dagger   \,,  \label{eq:ADC1Hl1} \\[1mm]
\gamma_{H\ell^{(3)}, H\Box} & =  \frac{g_2^2}{6}  - \frac{1}{2}  \hspace{0.25mm}  y_e y_e^\dagger \,.  \label{eq:ADC1Hl3}
\end{align} 
A sum over flavour indices is implicit in the above equations and the index $\psi$  in~(\ref{eq:C1fH}) and~(\ref{eq:ADC1fH})  can take the values $\psi = u, d, e$. Our results~(\ref{eq:C1fH}) to~(\ref{eq:C1Hd}) and~(\ref{eq:C1He}) to~(\ref{eq:C12y}) for the one-loop matching corrections of the fermionic dimension-six SMEFT operators agree with~(A.24) to (A.34) as given in~\cite{Jiang:2018pbd}. On the other hand, the correction~(\ref{eq:C1Hud}) was missed in~\cite{Jiang:2018pbd}. In addition, the anomalous dimension expressions (\ref{eq:ADC1fH}) to (\ref{eq:ADC1Hl3}) fulfil the one-loop SMEFT RG equations collected in~\cite{Jenkins:2013zja,Jenkins:2013wua,Alonso:2013hga}. Finally, note that the operator~$Q_{2y}$ is a linear combination of several four-fermion operators in the Warsaw basis~\cite{Wells:2015uba}. The~Wilson coefficient  $C_{2y} $ does not receive a logarithmic correction, since the (purely bosonic) tree-level operators $Q_{H\Box}$ and $Q_{H}$ obviously cannot mix into four-fermion operators  at one loop. 

\acknowledgments We thank the authors of \cite{Ellis:2017jns} and \cite{Jiang:2018pbd} for promptly confirming our results and for useful comments on the manuscript. We also thank Xiaochuan Lu for valuable feedback, and Jos{\'e} Santiago for making us aware of the ongoing work~\cite{matchmaker} and helpful discussions. We are grateful to Benjamin Summ for pointing out that the correction to the operator $Q_{Hud}$ was missing in previous versions of this work. The Feynman diagrams shown in this paper have been drawn with {\tt JaxoDraw}~\cite{Binosi:2008ig}. MR, EV and AW have been partially supported by the DFG Cluster of Excellence 2094 ORIGINS, the Collaborative Research Center SFB1258, the BMBF grant 05H18WOCA1 and thank the MIAPP for hospitality. MR is supported by the Studienstiftung des deutschen Volkes.

\appendix

\section{UOLEA results}
\label{sec:UOLEA}

In this appendix we apply functional methods to perform the one-loop matching and point out some pieces which were missed in the previous calculation~\cite{Ellis:2017jns}. Once these missing parts are accounted for, the results obtained in the UOLEA framework agree with the expressions of our diagrammatic calculation described in Section~\ref{sec:calculation}. We demonstrate this agreement explicitly for the contribution from heavy-particle loops to the one-loop matching corrections to the Wilson coefficients $C_{H \Box}$ and $C_H$. Our discussion follows the general line of reasoning presented in the articles~\cite{Henning:2014wua,Drozd:2015rsp,Ellis:2017jns} and we refer to the works~\cite{Henning:2016lyp,Ellis:2016enq,Fuentes-Martin:2016uol} for CDE and UOLEA formulations including heavy-light loops. 

Starting from the Lagrangian~(\ref{eq:lagrangianphi}) one can obtain the low-energy effective action by performing the functional integral over the $\phi$ field. The part originating from heavy particle loops   is given by
\beq \label{eq:Seff}
\begin{split}
S_{\rm eff} [H] & \simeq S[H,\phi_c]  + \frac{i}{2} \text{Tr}\ln\bigg(-\left.\frac{\delta^2 S}{\delta\phi^2}\right|_{\phi=\phi_c}\bigg) \\
& \simeq S[H,\phi_c]  +  \frac{i}{2} \text{Tr}\ln\left(-P^2 + M^2 + U_\phi \right ) \,.
\end{split}
\eeq
Here $\phi_c = \phi_c [H]$ is the solution of the classical equation of motion of $\phi$,~i.e.~$\delta S/\delta \phi \big |_{\phi=\phi_c} = 0$, $P_\mu = i\partial_\mu$ and~$U_\phi$  in the case of~(\ref{eq:lagrangianphi}) takes the form 
\beq \label{eq:Uphi}
U_\phi = \left.\frac{\partial^2 \mathcal{L}_\phi}{\partial \phi^2}\right|_{\phi=\phi_c} = \kappa |H|^2 + \mu \phi_c +\frac{1}{2} \lambda_\phi \phi_c^2\,.
\eeq
In order to find an expression for $\phi_c$ we perturbatively solve the equation of motion of~$\phi$, which explicitly reads
\beq \label{eq:EOMphi}
\left (\Box +M^2 + \kappa |H|^2 \right )\phi = -A |H|^2 - \frac{\mu}{2}\phi^2 -\frac{\lambda_\phi}{6}\phi^3\,. 
\eeq
Making the ansatz $\phi_c = \phi^{(0)}_c + \phi^{(1)}_c + \phi^{(2)}_c + \ldots$ with $\phi^{(k)}_c =  O(\mu^{l} \lambda_\phi^{n})$ and $k = l + n$ it is straightforward to obtain 
\begin{align}
\phi^{(0)}_c &= -\,\frac{1}{\Box + M^2 + \kappa |H|^2} \, A|H|^2 \,,\\[2mm]
\phi^{(1)}_c &= -\,\frac{1}{\Box + M^2 + \kappa |H|^2} \left ( \frac{\mu }{2} \, \big( \phi^{(0)}_c \big )^2+ \frac{\lambda_\phi}{6} \,  \big ( \phi^{(0)}_c \big )^3 \right ) \,,\\[2mm]
\phi^{(2)}_c &= -\,\frac{1}{\Box + M^2 + \kappa |H|^2} \left ( \mu \hspace{0.25mm} \phi^{(0)}_c \phi^{(1)}_c + \frac{\lambda_\phi}{2}   \big( \phi^{(0)}_c \big )^2  \phi^{(1)}_c \right ) \,.
\end{align}
Expanding $\phi_c$ up to four Higgs fields and two derivatives or six Higgs fields and no derivative, we then find the following expression
\beq \label{eq:phic}
\begin{split}
\phi_c = &-\frac{A}{M^2} |H|^2 + \bigg(\frac{A\kappa}{M^4}-\frac{A^2 \mu}{2M^6}\bigg) |H|^4 + \frac{A}{M^4} \Box |H|^2 \\[1mm]
& -\bigg(\frac{A \kappa}{M^6}- \frac{A^2 \mu}{M^8} \bigg)|H|^2 \Box  |H|^2 - \bigg(\frac{A\kappa^2}{M^6} -\frac{A^3 \lambda_\phi + 9A^2 \kappa \mu}{6M^8}  + \frac{A^3 \mu^2 }{2 M^{10}}\bigg) |H|^6 \,.
\end{split}
\eeq
Comparing the above result for $\phi_c$ to~(4.2) of~\cite{Ellis:2017jns} one observes that while the first three terms of~(\ref{eq:phic}) agree with the  $|H|^2$, $|H|^4$ and $\Box |H|^2$ contributions given in the latter work, the $|H|^2 \Box  |H|^2$ and $ |H|^6$ contain additional pieces, all of which vanish in the limit  $\mu \to 0$. 

These additional terms affect the matching contributions from heavy loops to the one-loop Wilson coefficients $C_{H \Box}^{(1)}$ and $C_H^{(1)}$, which consequently differ from the results presented in the work~\cite{Ellis:2017jns}. Considering the full solution of the classical  equation of motion, we find that the heavy-loop contribution to $C_{H \Box}^{(1)}$ is given by
\beq \label{eq:CHBox1heavy}
\begin{split}
C_{H \Box}^{(1)} \Big|_{\rm heavy} & =  -\left ( \frac{A^2\lambda_\phi +  A \kappa \mu}{2 M^6} - \frac{A^2  \mu^2}{2 M^8} \right ) \tilde f_2  \\[1mm]  
& \phantom{xx} + \left ( \frac{A \kappa \mu}{M^4} - \frac{A^2  \mu^2}{M^6} \right )  \tilde f_4  + \left ( \frac{\kappa^2}{2} - \frac{A \kappa \mu}{M^2} + \frac{A^2 \mu^2}{2 M^4} \right ) \tilde f_7  \\[2mm]
& =  -\,\frac{\kappa^2}{24 M^2} - \frac{6 A^2 \lambda_\phi  +  5 A \kappa \mu}{12 M^4} + \frac{11 A^2 \mu^2}{24 M^6} -\frac{A^2 \lambda_\phi}{M^4}   \hspace{0.5mm} \ln \frac{\mu_M}{M} \,,
\end{split}
\eeq
where to obtain the final result we have inserted the expressions for the universal coefficients~$\tilde f_N$ reported in Appendix~B of~\cite{Ellis:2017jns}. Notice that only the prefactor of $\tilde f_2$ in~(\ref{eq:CHBox1heavy})  differs  from the result~(4.10) presented in the article~\cite{Ellis:2017jns}.  Diagrammatically the observed difference of $A^2 \mu^2 \left ( 1 + 2  \ln \mu_M/M  \right )/(2 M^6)$ is due to propagator-type tadpole contributions --- see~the last  diagram in Figure~\ref{fig:heavyhbox} --- that have effectively been missed in the latter calculation. 

In the case of the heavy one-loop matching contributions to the Wilson coefficient of the operator~$Q_H$ $\big($cf.~(\ref{eq:lagrangianSMEFT}) and~(\ref{eq:warsaw})$\big)$, we instead obtain the following expression 
\beq \label{eq:CH1heavy}
\begin{split}
C_H^{(1)} \Big|_{\rm heavy} & =  -\left( \frac{A^2 \kappa  \lambda_\phi + A \kappa ^2 \mu}{2 M^6} -\frac{4 A^3 \mu  \lambda_\phi + 9 A^2 \kappa \mu ^2}{12 M^8} + \frac{A^3 \mu ^3}{4 M^{10}} \right) \tilde f_2  \\[1mm]
& \phantom{xx} + \left(\frac{A^2 \kappa  \lambda_\phi +2 A \kappa ^2 \mu}{2 M^4}- \frac{A^3 \mu \lambda_\phi +3 A^2 \kappa  \mu ^2}{2 M^6} + \frac{A^3 \mu ^3}{2 M^8} \right) \tilde f_4  \\[1mm]
& \phantom{xx} + \left(\frac{\kappa ^3}{2} -\frac{3 A \kappa ^2 \mu}{2 M^2} +\frac{3 A^2 \kappa  \mu ^2}{2 M^4} -\frac{A^3 \mu ^3}{2 M^6} \right)\tilde f_8  \\[2mm] 
& = -\,\frac{\kappa ^3}{12 M^2} - \frac{2 A^2 \kappa  \lambda_\phi + A \kappa ^2 \mu}{4 M^4}  + \frac{2 A^3 \mu  \lambda_\phi + 3 A^2 \kappa \mu ^2}{6 M^6} - \frac{A^3 \mu ^3}{6 M^8} \\[1mm]
& \phantom{xx} - \left (\frac{A^2\kappa \lambda_\phi }{2M^4} - \frac{A^3\mu \lambda_\phi }{6 M^6} \right) \ln \frac{\mu_M}{M}  \,.
\end{split}
\eeq
Apart from the prefactor of $\tilde f_2$, the latter result agrees with~(4.9) of~\cite{Ellis:2017jns}. The resulting difference of $A^2 \mu^2 \left ( A \mu - 2 M^2 \kappa \right ) \left ( 1  + 2 \ln \mu_M/M \right )/(4 M^8)$ can again be traced back to  propagator-type tadpole contributions that have not been correctly included in the latter article. Our formula~(\ref{eq:CH1heavy}) is in accord with the preliminary results presented in the talk~\cite{matchmaker}, where small discrepancies with the formula for $C_{H}^{(1)}$ given in~\cite{Ellis:2017jns} were already observed.

The  heavy-light contributions to the one-loop matching corrections to the dimension-six SMEFT operators $Q_{H \Box}$ and $Q_H$ are not affected by the additional terms in~(\ref{eq:phic}). In fact, the operators generated by heavy-light loops that are proportional to $\phi_c$ appear with at least two additional Higgs fields compared to $Q_{H \Box}$ and $Q_H$. As a result the missing terms only affect the one-loop matching of operators with a mass dimension of eight or higher. However, when trying to reproduce the results in~\cite{Ellis:2017jns} we discovered an unrelated typo in the heavy-light contribution to $C_H^{(1)}$ that appears in the prefactor of $\tilde{f}_{4A}$ in~(4.8) of that work. We find that the actual contribution of the universal  coefficient $\tilde{f}_{4A}$ to the Wilson coefficient $C_H^{(1)}$ should read 
\beq \label{eq:CHloopf4A}
C_H^{(1)} \Big|_{\tilde{f}_{4A}} =  \frac{3A^2 \kappa^2}{M^4}-\frac{A^3 \kappa \mu}{M^6} \,. 
\eeq
Once the additional  terms in~(\ref{eq:phic})  and the correction~(\ref{eq:CHloopf4A})  are taken into account we recover the diagrammatic results for  $C_{H \Box}^{(1)}$ and~$C_{H}^{(1)}$ originating from heavy and heavy-light pure scalar loops, meaning that our UOLEA calculation reproduced the terms in the first two lines of~(\ref{eq:CHBoxloop}) as well as the terms  in the first three lines of~(\ref{eq:CHloop}). Note that the above mistakes and typos are also present in~(3.1) and~(3.2) of~\cite{Jiang:2018pbd}, which employed the UOLEA master formulas of~\cite{Ellis:2017jns} to obtain the aforementioned results. 

\section{RG evolution of SSM parameters}
\label{sec:SSMRG}

We define the one-loop anomalous dimensions $\gamma_{x}$ that enter the RG evolution of the  parameters $x$ by
\beq \label{eq:RG}
\frac{d x}{d \ln \mu_R}  =  \frac{\gamma_{x}}{(4 \pi)^2} \,.
\eeq 
Renormalising all UV poles including those arising from $\phi$ tadpoles in the  $\overline{\rm MS}$ scheme, the anomalous dimensions of the parameters $M^2$, $A$, $\kappa$, $\mu$, $\lambda_\phi$ and $\lambda_h$  read
\begin{align}
\gamma_{M^2} & = M^2 \lambda_\phi + 4 A^2 \,, \label{eq:gammmaM2} \\[1mm]
\gamma_A &  = A \left ( 4 \kappa + 6 \lambda_h - \frac{3}{2} \left (g_1^2 + 3 g_2^2 \right )+2 y_2 \right )   \,,  \label{eq:gammmaA} \\[1mm]
\gamma_\kappa &  = \kappa \left ( 4 \kappa + \lambda_\phi + 6 \lambda_h - \frac{3}{2} \left (g_1^2 + 3 g_2^2 \right ) +2  y_2 \right )  \,, \label{eq:gammmakappa}    \\[1mm]
\gamma_\mu &  =  2\mu \lambda_\phi + 12 A \kappa \,, \label{eq:gammmamu} \\[1mm]
\gamma_{\lambda_\phi} &  =  12  \kappa^2 + 3  \lambda_\phi^2 \,, \label{eq:gammmalphi} \\[1mm]
\gamma_{\lambda_h} &  =   \kappa^2+ 12 \lambda_h^2  -3 \lambda_h \left (g_1^2 + 3 g_2^2 \right ) + \frac{3}{4} \left ( g_1^4 + 2 g_1^2 g_2^2 + 3 g_2^4 \right ) + 4 \lambda_h y_2 - 4 y_4 \,, \label{eq:gammmalh} 
\end{align}
with
\begin{align} 
y_2 & = {\rm Tr} \left ( 3 \hspace{0.125mm} y_u^\dagger  y_u +  3 \hspace{0.125mm}  y_d^\dagger  y_d  + y_e^\dagger y_e \right ) \,, \label{eq:y2} \\[1mm]
y_4 & = {\rm Tr} \left ( 3 \hspace{0.125mm}  y_u^\dagger  y_u y_u^\dagger  y_u  + 3 \hspace{0.125mm} y_d^\dagger  y_d  y_d^\dagger  y_d  + y_e^\dagger y_e  y_e^\dagger y_e \right ) \,. \label{eq:y4}
\end{align}
Examples of Feynman graphs that contribute to the anomalous dimensions $\gamma_{M^2}$, $\gamma_A$, $\gamma_\kappa$ and  $\gamma_\mu$ are shown on the right of Figure~\ref{fig:wfr} as well as in Figure~\ref{fig:phiprop} to Figure~\ref{fig:mu}. We add that the results for $\gamma_{\lambda_\phi}$ and $\gamma_{\lambda_h}$ are not needed in the context of this work, but we provide them for completeness.

In Section~\ref{sec:oneloopmatching} of this article we have presented our final results~(\ref{eq:CHBoxloop}) and~(\ref{eq:CHloop})  for the one-loop matching corrections $C_{H \Box}^{(1)}$ and $C_{H}^{(1)}$. In both cases we have observed that the logarithmic corrections to the Wilson coefficients involve only anomalous dimensions that depend only on SM couplings but not on SSM parameters. Below we explicitly show how this feature arises.  The given formulae should also facilitate a comparison to the existing computations~\cite{Ellis:2017jns,Jiang:2018pbd} as well as to the preliminary results presented in the talk~\cite{matchmaker}.

In order to derive the logarithmic terms that arise from the renormalisation of the SSM parameters, we first notice that if the tree-level Wilson coefficients $C_k^{(0)}$ are expressed through the SSM parameters $x$ one has 
\beq \label{eq:dC0dln}
\frac{d C_k^{(0)}}{d \ln \mu_R} = \sum_x \frac{\partial C_k^{(0)}}{\partial x} \frac{d x}{d \ln \mu_R} = \frac{1}{(4 \pi)^2}  \sum_x  \frac{\partial C_k^{(0)}}{\partial x} \hspace{0.25mm} \gamma_x \,,
\eeq   
where the sum over $x$ includes $M^2$, $A$, $\kappa$, $\mu$, $\lambda_\phi$ and $\lambda_h$, and  in the last step we have used the definition~(\ref{eq:RG}). Integrating~(\ref{eq:dC0dln}) from $\mu_M$ to $M$ then gives  rise to the  logarithmic corrections that are associated to the renormalisation of the SSM parameters.

Applying the master formula~(\ref{eq:dC0dln}) to the case of the Wilson coefficient $C_{H \Box}^{(1)}$, we find the following logarithmic terms
\beq 
\begin{split} \label{eq:CHBoxloopLL}
C_{H \Box}^{(1)} \Big|_{\ln \frac{\mu_M}{M}} & = \frac{4 A^2 \kappa - A^2 \lambda_\phi}{M^4} + \frac{2A^4}{M^6} - \frac{5 A^2 \left (g_1^2 + 3 g_2^2 \right )}{6 M^4}  + \frac{A^2 \gamma_{M^2}}{M^6} -  \frac{A \gamma_{A}}{M^4} \\[2mm] 
& = \left ( 12 \lambda - \frac{4}{3} \left (g_1^2 + 3 g_2^2 \right ) + 4 y_2 \right ) \left( -\frac{A^2}{2 M^4} \right ) \,. 
\end{split}
\eeq
The  first three terms in the first line of~(\ref{eq:CHBoxloopLL}) result from the heavy and heavy-light loop diagrams shown in Figure~\ref{fig:heavyhbox} to Figure~\ref{fig:heavylightgaugehbox}. The terms proportional to $\gamma_{M^2}$ and $\gamma_A$ instead arise from the renormalisation of the parameters $M^2$ and $A$ that enter the tree-level Wilson coefficient~(\ref{eq:CHBoxtree}). In the second line of~(\ref{eq:CHBoxloopLL}) we can manifestly see that there is a cancellation of logarithmic terms involving the combinations of SSM parameters that do not form a SMEFT Wilson coefficient, yielding the logarithmic correction quoted in~(\ref{eq:CHBoxloop}) as final result. 

In the case of the Wilson coefficient $C_H^{(1)}$ using~(\ref{eq:dC0dln}) instead leads to
\beq
\begin{split} \label{eq:CHloopLL}
C_H^{(1)} \Big|_{\ln \frac{\mu_M}{M}} & =  -\,\frac{A^2 \kappa \lambda_\phi+36 A^2 \kappa \lambda-12 A^2 \kappa^2-40A^2 \lambda^2}{2 M^4} \\[1mm] 
& \phantom{xx} + \frac{18 A^4 \kappa -12 A^3 \kappa \mu + A^3 \mu \lambda_\phi + 36 A^3 \lambda \mu}{6 M^6}  - \frac{A^5 \mu}{M^8}  - \frac{10 A^2 \lambda g_2^2}{3 M^4}  \\[1mm] 
& \phantom{xx} + \left ( \frac{A^2 \kappa}{M^6}  - \frac{A^3 \mu}{2 M^8} \right ) \gamma_{M^2}  - \left ( \frac{A \kappa}{M^4}  - \frac{A^2 \mu}{2 M^6} \right ) \gamma_A - \frac{A^2 \gamma_\kappa}{2M^4} +  \frac{A^3 \gamma_\mu}{6 M^6} \\[2mm]
& = \left ( -40 \lambda^2 + \frac{20 \lambda g_2^2}{3}  \right )  \left( -\frac{A^2}{2 M^4} \right )  \\[1mm]
& \phantom{xx} + \left ( 54 \lambda - \frac{9}{2} \left (g_1^2 + 3 g_2^2 \right ) + 6 y_2 \right ) \left ( \frac{A^3 \mu}{6 M^6} -\frac{A^2 \kappa}{2M^4}  \right ) \,. 
\end{split}
\eeq
The  first two lines of the above expression correspond to the contributions from heavy and heavy-light graphs, while  the terms proportional to the anomalous dimensions $\gamma_{M^2}$, $\gamma_A$,  $\gamma_\kappa$ and  $\gamma_\mu$ are the counterterm contributions that are associated to the renormalisation of the relevant SSM parameters appearing in the tree-level Wilson coefficient $C_H^{(0)}$. Notice that the final result in~(\ref{eq:CHloopLL}) agrees with the logarithmic correction that we have obtained in~(\ref{eq:CHloop}), and that these terms have the correct form to allow for a resummation of large logarithms using the RG equations of the dimension-six SMEFT operators $Q_{H \Box}$ and $Q_H$ derived in~\cite{Jenkins:2013zja,Jenkins:2013wua,Alonso:2013hga}. 

%\bibliographystyle{apsrevmod}
%\bibliography{SSM}


\end{document}